\documentclass[aps,pra,showkeys,a4paper,groupedaddress,floatfix]{revtex4}
\usepackage{graphicx}
\usepackage{tabularx}
\usepackage{amsmath}
\usepackage{amsfonts}
\usepackage{amssymb}
\usepackage{natbib}
\begin{document}
\title{Transient behavior of surface plasmon polaritons scattered at a subwavelength groove}
\date{\today}
\author{G. L\'ev\^eque}\author{O. J. F. Martin}\affiliation{Laboratoire de Nanophotonique et M\'etrologie\\Ecole Polytechnique
F\'ed\'erale de Lausanne\\CH1015-Lausanne, Switzerland}\email{gaetan.leveque@epfl.ch}
\author{J. Weiner}\affiliation{IRSAMC/LCAR\\
Universit\'e Paul Sabatier, 118 route de Narbonne,\\31062
Toulouse, France}\email{jweiner@irsamc.ups-tlse.fr}


\begin{abstract}
We present a numerical study and analytical model of the optical
near-field diffracted in the vicinity of subwavelength grooves
milled in silver surfaces. The Green's tensor approach permits
computation of the phase and amplitude dependence of the diffracted
wave as a function of the groove geometry. It is shown that
the field diffracted along the interface by the groove is equivalent
to replacing the groove by an oscillating dipolar line source.
An analytic expression is derived from the Green's function
formalism, that reproduces well the asymptotic surface plasmon
polariton (SPP) wave as well as the transient surface wave in the
near-zone close to the groove. The agreement between this model and the
full simulation is very good, showing that the transient ``near-zone"
regime does not depend on the precise shape of the groove. Finally,
it is shown that a composite diffractive evanescent wave model
that includes the asymptotic SPP can describe the wavelength evolution
in this transient near-zone. Such a semi-analytical model may be useful
for the design and optimization of more elaborate photonic circuits
whose behavior in large part will be controlled by surface waves.
\end{abstract}

\maketitle

\section{Introduction}

The analysis of light diffraction and transmission through a slit
has a long history in physical optics\,\cite{B54,BW80}.  As
discussed by Kowarz\,\cite{KowarzThesis,K95} the analysis can be
separated into two problems: the boundary value problem and the
propagation problem. The boundary value problem concerns the
determination of the field immediately at the output plane, and
interest is usually concentrated on the boundary values in the
vicinity of the slit. The propagation problem involves
determination of the field at points in the halfspace beyond the
output plane in the near- and far-field.  Recent interest in light
transmission through subwavelength periodic structures with
subwavelength pitch\,\cite{ELG98,BDE03} has stimulated some
experimental\,\cite{LT04,GAV06a,GAV06b,GAW07,KGA07} and many
theoretical
studies\,\cite{T99,T02,CL02,LSH03,GLE03,LHR05,XZM04,XZM05,LH06}
with the aim of better understanding the nature of the light field
at the surface (the boundary value problem) and the influence this
surface field on light transmission (the propagation problem).

We report here numerical simulations of single groove and
slit-groove structures using a Green's tensor method to solve the
Maxwell field equations near the subwavelength structures on the
metal/free-space interface.  The simulations are compared with
recent experimental results on single slit-groove
structures\,\cite{GAV06a,GAV06b,GAW07} and essentially confirm the
observed amplitude and phase evolution of the scattered waves as a
function of groove geometry and groove-slit distance.  We then
show that the results of the full numerical simulation can be
recovered by replacing the groove structure with a simple
oscillating dipole source and again applying the Green's tensor
method to find the near- and far-field along the surface.  This
oscillating dipole picture is consistent with recent charge and
field distributions at metal-slit and metal-groove boundaries
found numerically by a finite-difference time domain (FDTD)
technique\,\cite{XZM04} and permits calculation of both the
propagating and evanescent contributions to the scattered field.
Finally we also present a simple analytic
aperture-in-opaque-screen model, in the same vein as earlier
models\,\cite{KowarzThesis,K95,LT04}, but with a boundary
condition that posits the SPP mode at the metal/free-space
interface.  Comparison of the Green's tensor simulations to
the analytic model helps to physically interpret the numeric
results in terms of surface-wave modes.  The oscillating
dipole picture, however, provides deeper insight into the essential physics
of surface wave generation at the groove while overcoming the
limitations of any fixed-boundary-condition model.

\section{Numerical simulations\label{section-numerical simulations}}
The numerical simulations are performed with the Green's tensor
method\,\cite{MGD95,PM01}. This method is very convenient for the
study of finite-size, two-dimensional (2D) or three-dimensional
(3D) objects embedded in a multilayered background.  It relies on
the resolution of the Lippmann-Schwinger equation of the electric
field.
\begin{equation}
\mathbf{E(r)}=\mathbf{E}^0(\mathbf{r})+k_0^2\int_Vd\mathbf{r^{\prime}}\mathbf{G}(\mathbf{r,r^{\prime}})
\Delta\epsilon(\mathbf{r^{\prime}})\mathbf{E(r^{\prime})},
\end{equation}
where $\mathbf{G}(\mathbf{r,r^{\prime}})$ is the Green's tensor
associated with the stratified background,
$k_0^2=\omega^2\epsilon_0\mu_0$ is the square of the wave
propagation parameter, and $\Delta\epsilon(\mathbf{r})$ is the
``dielectric contrast," the relative permittivity difference
between the scatterer of volume $V$ and the adjacent layer.  The
Green's tensor itself is the solution of a vector wave equation
with a point dipole source,
\begin{equation}
\boldsymbol{\nabla}\times\boldsymbol{\nabla}\times\mathbf{G}(\mathbf{r,r^{\prime}})-k_0^2\epsilon\mathbf{G}
(\mathbf{r,r^{\prime}})=\mathbf{1}\delta(\mathbf{r}-\mathbf{r^{\prime}}),
\end{equation} where $\mathbf{1}$ is the unit tensor. The 3D Green's
tensor represents the electric field at $\mathbf{r}$ produced by
three orthogonal unit dipoles located at $\mathbf{r^{\prime}}$ in
a layer of dielectric constant $\epsilon$. An advantageous
distinguishing property of the Green's tensor method is that only
the objects of interest need be discretized. The boundary
conditions at infinity are included in the Green's tensor of the
multilayered background. In the present case, the calculation
takes a very short time because of the small size (some tens of
nanometers) of the groove and slit. Details and extensive
references concerning this method can be found in a recent
review\,\cite{G05}.

\begin{figure}
\includegraphics[width=8cm]{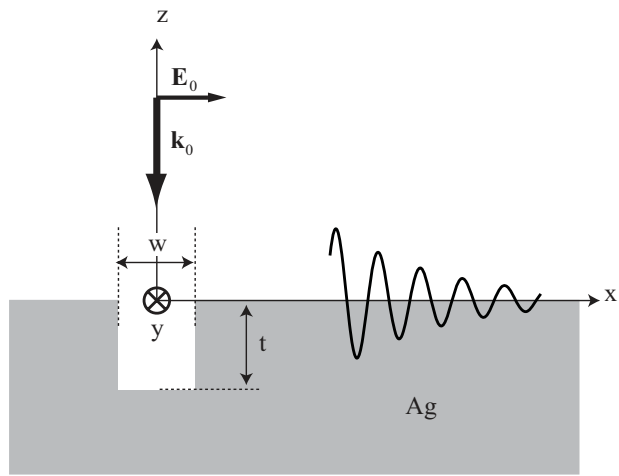}
\caption{\label{fig1}Geometry of the single groove system.}
\end{figure}

The system initially studied is shown in Fig.\,\ref{fig1}.  It is
an empty groove milled into a metallic substrate with the groove
profile along $x$ and extending invariantly along $y$. Outside the
groove, the metal/free-space interface lies in the $0xy$ plane
with $z$ axis in the vertical direction. The rectangular section
of the groove is characterized by width $w$ and depth $t$. The
incident free-space electromagnetic plane wave, with E-field
polarized along $x$, impinges on the groove and interface at
normal incidence. Because of the reduced dimensionality of the
problem, all scattered waves, propagating and evanescent, are
restricted to the $0xz$ plane.

\begin{figure}
\begin{center}
\includegraphics[width=8cm]{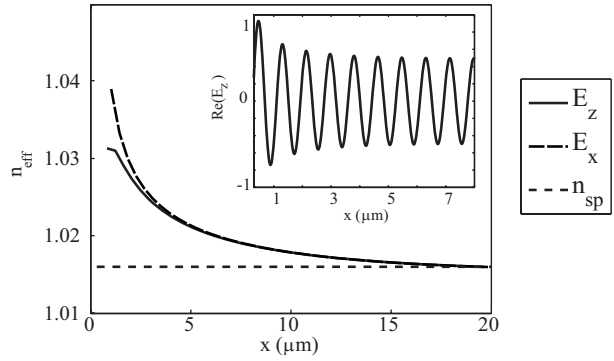}
\caption{\label{figcomp}Evolution of the $E_z$ (solid curve) and
$E_x$ (dashed curve) effective index
$n_{\mathrm{eff}}(x)=\lambda(x)/\lambda_0$ with the distance from
the center of the groove. The groove dimensions are: $t=100$ nm,
$w=100$ nm. Inset: evolution of the real part of $E_z$ with $x$,
along the $z=0^+$ line.}
\end{center}
\end{figure}

The electric field is calculated along the line $z=0^+$, $x>0$.
All simulations are performed at the free-space wavelength
$\lambda_0=852$~nm. In order to compare these results with
previously published studies\,\cite{GAV06a,GAW07,LH06}, the
relative permittivity (dielectric constant) of silver is taken to
be $\epsilon=-33.22+1.17\,i$. The corresponding propagation length
against absorption is $680~\mu$m, quite long compared to $\lambda_0$, because the imaginary
term in the dielectric constant is small for silver at this
wavelength.  The results of the calculation are plotted in
Fig.\,\ref{figcomp}.
The principal plot in Fig.\,\ref{figcomp} shows the evolution of
the effective index $n_{\mathrm{eff}}=\lambda(x)/\lambda_0$ as a
function of the distance from the center of the groove, for both
$E_x$ and $E_z$ components. The dotted line indicates the
effective index of the SPP guided wave,
\[n_{\mathrm{spp}}=\Re\left[\left(\frac{1+\epsilon_m}{\epsilon_m}\right)^{-1/2}\right]=1.0154
\]
for the silver/free-space interface. The variation of the
effective index is sightly different for the two components close to the groove, but both
curves converge rapidly after 3~$\mu$m. In both cases, the
effective index is larger than $n_{\mathrm{spp}}$ out to $\sim
10~\mu$m, but for greater distances, it converges to the expected
$n_{\mathrm{spp}}$.  The results of the simulation are consistent
with the measurements of Ref.\,\cite{GAV06a}, that reported a
value of $n_{\mathrm{surf}}=1.04 \pm 0.01$ over a distance of
$\sim 6~\mu$m.  They are also consistent with recent
finite-difference time-domain (FDTD) simulations on silver
surfaces\,\cite{GAW07} as well as similar measurements and
simulations on gold surfaces\,\cite{KGA07}.  The inset of
Fig.\,\ref{figcomp} shows the $z$-component of the electric field
diffracted by the groove along the interface. We can clearly
identify two regimes. The first extends out to $\simeq 3~\mu$m
along $x$ and is characterized by a relatively rapid decay of the
amplitude. For further distances, the amplitude decreases much
more slowly (due to absorption) and appears constant over the
displayed range. This two-step evolution is characteristic of a
transient regime within the first few micrometers from the groove.
Since the incident wave is TM polarized (E-field parallel to $x$),
$E_z$ belongs only to the scattered field, and does not interfere
with the incident wave. For this reason, the mean value of the
real part of $E_z$ in the \emph{total} field along the interface
must be zero. This is not the case for the $x$-component, as the
incident field is polarized along the $x$ direction.

\begin{figure}[htb]
\centering\includegraphics[width=8cm]{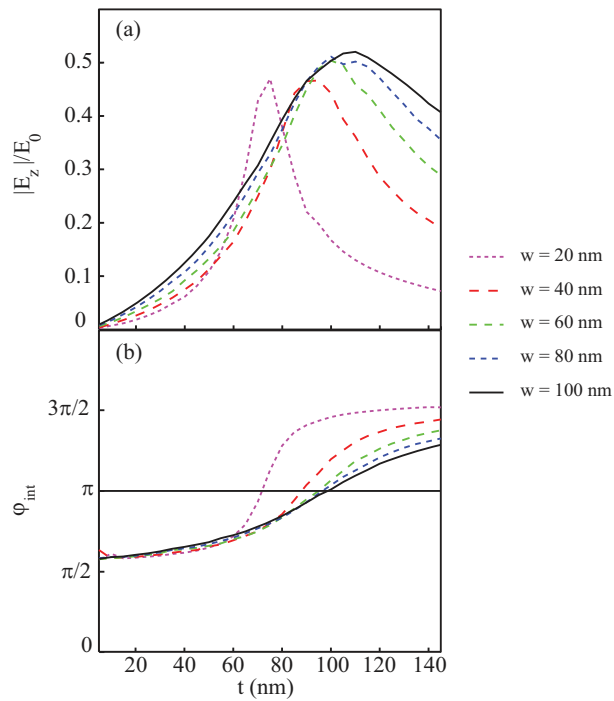}
\caption{(a) Evolution of the amplitude of the
$z$--component of the electric field, 18~$\mu$m away from the
groove ; (b) Evolution of the intrinsic phase $\varphi_{\mathrm{int}}$ of the
$z$--component of the electric field diffracted by the groove (color online). }
\label{ampphase}\end{figure}

\begin{figure}[htb]
\centering\includegraphics[width=8cm]{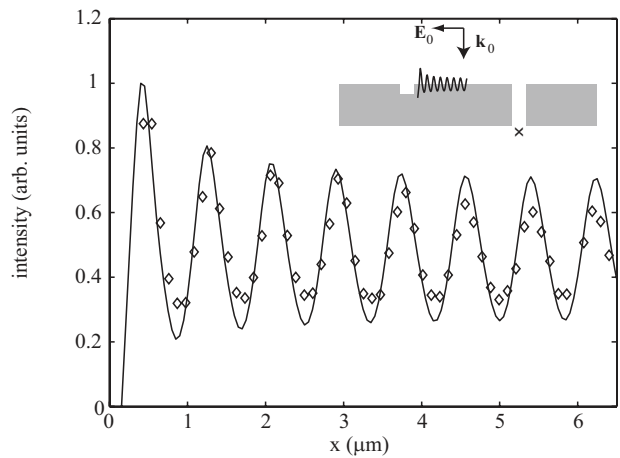}
\caption{Comparison between the experimental results of
Ref\,\cite{GAV06a} and the numerical simulations performed with a
groove of width $w=100$ and depth of $t=120$ nm. The silver slab
is 400-nm thick, and the slit 100-nm wide. The electric-field
intensity was computed at the exit of the slit (black cross).}
\label{figexp}\end{figure}

The simulations also show that the phase and the amplitude of the
diffracted wave are sensitive to the groove dimensions, as
reported in experiments\,\cite{GAV06b}. In Fig.~\ref{ampphase}(a) and (b) are plotted the amplitude and phase evolution of the
surface wave, at large distance (18 $\mu$m).  The phase
$\varphi_{\mathrm{int}}$ has been determined by comparison with a
cosine function representing the SPP guided surface wave. It
corresponds to the asymptotic phase of the diffracted wave, called
the ``intrinsic phase" in Ref.\,\cite{GAV06b}.  The evolution of
the scattered wave phase and amplitude as a function of groove
depth is typical of a resonant phenomenon. Here, the resonance
concerns standing-wave modes created inside the groove. The $w$
and $t$ dimensions of the groove may be varied so as to produce a
cavity that resonates when excited by an incident surface wave. At
resonance a vertical standing mode dominates the field
distribution inside the groove.  Because of the boundary
conditions, the electric field must be almost null at the bottom
of the groove.  Near a resonance, the phase of the diffracted wave
varies rapidly and passes through an inflection point, while the
amplitude reaches a maximum. As can be observed in
Fig.\,\ref{ampphase}, the amplitude is maximal when the phase is
almost $\pi$. The experimental results\,\cite{GAV06b} reported an
intrinsic phase value of $\pi/2$ at the resonance groove depth.
This difference of $\pi/2$ between the experiment and the
simulation arises from the fact that $x$-- and $z$-- components of
the surface wave E-field oscillate in quadrature.  The experiment
essentially measures the intrinsic phase difference in a far-field
interference pattern between two oscillating dipoles oriented
along $x$: one localized at the corners of a slit and the other
localized at the groove (see inset of Fig.~\ref{figexp}).  Thus
the experiment is sensitive to the intrinsic phase of the
$x$--component of the surface E-field while the simulation
calculates the intrinsic phase of the $z$--component.  After
taking this quadrature phase difference into account, we see that
the simulations are consistent with the measurements.
 The groove depth for which the
resonance occurs increases with the width. For a width of
$w=100$~nm, the simulations yield an optimal depth $t\approx
110$~nm. For shallow depths, some tens of nanometers, the
amplitude and phase depend weakly on the width. At greater depths,
the amplitude becomes quite sensitive to the groove width, but the
phase does not change dramatically for widths $w>40$~nm.  Clearly
the phase and the amplitude of the diffracted wave are very
sensitive to the groove geometry. As the absolute groove depth is
difficult to determine experimentally, the simulations results for
ideal geometries may differ somewhat from the nominal parameters
of fabricated structures.

In the case of the slit-groove experiments reported in
\cite{GAV06a} the Green's tensor simulations produced the best
agreement with the experimental points by considering a depth of
$t=120$~nm, rather than the nominal experimental depth of 100~nm.
A comparison between the simulation and measurement is plotted in
Fig.~\ref{figexp}; only the initial amplitude has been normalized
to the experimental curve.  Although the experimental intensity
derives from a far-field interference fringe and the simulated
field is evaluated at the output-side plane, it is legitimate to
compare the two curves because the far-field signal is
proportional to the calculated field intensity at the output-side
slit exit. We note that the same slit-groove calculation performed
with $t=100$~nm is in excellent agreement with the simulation of
Lalanne et al.~\cite{LH06} using an entirely different simulation
technique.

\section{Field scattered by a dipole along the interface}\label{secdip}
\begin{figure}[htb]
\centering\includegraphics[width=8cm]{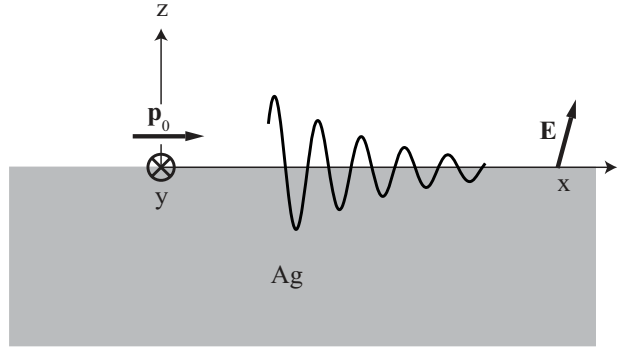} \caption{The
source is a dipole line located along the $y$ axis at $z=0^+$, and
oriented parallel to the $x$ axis. The field is computed along
$(x,z=0^+)$.} \label{fig13}\end{figure}

In this section we consider the 2D field radiated by a line dipole
$\mathbf{p}_0$ (rather than the 3D field radiated by a point
dipole) located just above the metal/free-space interface, as
indicated in Fig.\,\ref{fig13}. This approach has been applied
by Lalanne and Hugonin in\,\cite{LH06} to study the amplitude evolution
of the scattered magnetic field. The choice of placing the dipole
just above the surface may seem arbitrary, but it is shown in the
appendix that placing the dipole just under the interface leads to
the same conclusions. The dipole is aligned parallel to the
$x$-axis, consistent with the previous groove calculations of
section \ref{section-numerical simulations}. For the same reasons
discussed there, we only calculate the expression of the $z$
component of the electric field. The dipole oscillation wavelength
is 852~nm. We will use the Green's tensor formalism to extract a
simple expression for the field just above the interface.

The field radiated by the dipole at a point just above the
interface  at a distance $x$ is given by the equation:
\begin{equation}
\mbox{\bf E}(\mbox{\bf r},\mbox{\bf r}_0) = \frac{k_0^2}{\epsilon_0}\left[\mbox{\bf
G}_0(\mbox{\bf r}, \mbox{\bf r}_0)+\mbox{\bf G}_S(\mbox{\bf
r},\mbox{\bf r}_0)\right]\mathbf{p}_0,\nonumber
\end{equation}
with $\mathbf{r}(x,z)=(x,0^+)$ and $\mathbf{r}_0=(0,0^+)$.  We
denote the couple $(\mathbf{r},\mathbf{r_0})$ as $(x,0)$. The
$3\times 3$ tensors $\mathbf{G}_0$ and $\mathbf{G}_S$ are the
dyadic Green's functions associated with free space and the
metal/free-space surface at the considered wavelength. Thus, the
first term represents the field directly radiated by the dipole to
the observation point through free space. The second term
represents the field radiated to the observation point \emph{after
reflection from the surface}. The observation point is displaced
along $x$, on a line just above and parallel to the surface,
running through the dipole.  Due to symmetry of the dipole
radiation pattern, the $z$-component of the directly radiated term
along the line of observation points is 0, and we have:
\begin{eqnarray}
E_z(x)&=&\frac{k_0^2}{\epsilon_0}\left[\mbox{\bf G}_0(x,0)\mbox{\bf p}_0+\mbox{\bf G}_S(x,0)\mbox{\bf p}_0\right]
\cdot\mbox{\bf e}_z\nonumber\\
&=&\frac{k_0^2}{\epsilon_0}\left[\mbox{\bf G}_S(x,0)\mbox{\bf p}_0\right]\cdot\mbox{\bf e}_z\nonumber\\
&=&\frac{k_0^2}{\epsilon_0}G_S^{zx}(x,0)p_0,\label{eq3}
\end{eqnarray}
where $\mbox{\bf e}_z$ is the unit vector of the $z$ axis and
$G_S^{zx}$ is the $zx$ component of the surface Green's function,
i.e. the z-component of the Green's function produced by a dipole
aligned parallel to $x$.

Although there exist approximate expressions for $x$ small
compared to the wavelength (electrostatic approximation), these
expressions are not appropriate here because the line of
observation points extends far beyond a wavelength.  The exact
expression of the surface Green's function cannot be written in
closed form in direct space, but we can find an expression
susceptible to numerical evaluation by standard methods. The
Green's tensor is analytically defined in the half Fourier space
$(q,z)$, where $q$ is the spatial frequency parallel to the $x$
axis. The general expression can be found in reference
\cite{PM01}. The $G_S^{zx}$ component is given by:
\begin{equation}
G_S^{zx}(\mbox{\bf r},\mbox{\bf r}')=-\frac{i}{4\pi k_0^2}\int_{-\infty}^{+\infty} dq\,q\,R(q) e^{iqx},\nonumber
\end{equation}
where $R$ is the Fresnel reflection coefficient for TM (transverse
magnetic) polarization:
\begin{equation}
R(q)=\frac{\sqrt{\epsilon(\omega)k_0^2-q^2}-\epsilon(\omega)
\sqrt{k_0^2-q^2}}{\sqrt{\epsilon(\omega)k_0^2-q^2}+\epsilon(\omega)\sqrt{k_0^2-q^2}}.\label{R-de-q}
\end{equation}
Then, Eq.~(\ref{eq3}) becomes:
\begin{equation}\label{dirac}
E_z(x)=-\frac{i p_0}{4\pi \epsilon_0}\int_{-\infty}^{+\infty} dq\,q\,R(q) e^{iqx}.
\end{equation}
\begin{figure}[htb]
\centering\includegraphics[width=8cm]{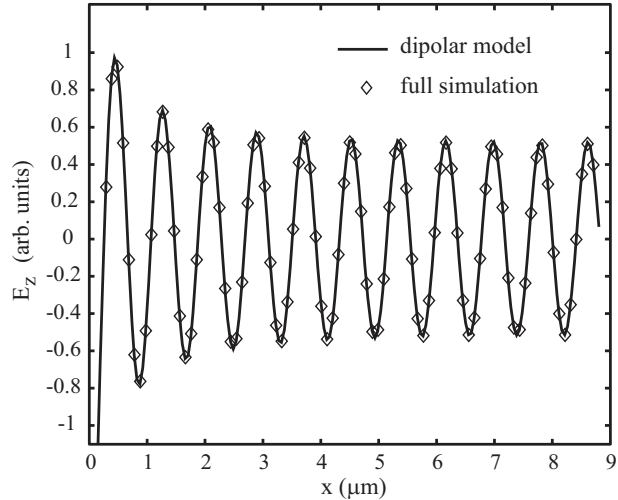}
\caption{Comparison between the field obtained from
Eq.~(\ref{dirac}) and with the full Green's tensor numeric
simulation.} \label{fig8}\end{figure} In order to interpret this
last equation, consider the value of the integral without the
reflection coefficient:
\begin{equation}
I(x)=\int_{-\infty}^{+\infty} dq\,q\,e^{iqx}.\nonumber
\end{equation}
This function $I$ is the derivative of the function $J$:
\begin{equation}
J(x)=-i\int_{-\infty}^{+\infty} dq\, e^{iqx},\nonumber
\end{equation}
proportional to a Dirac delta function located at $x=0$. So $I$ is
the derivative of a Dirac delta function, that in fact represents
a point dipole located at $x=0$: the $z$ component of the electric field in the direction of the dipole is 0 everywhere, except at two points infinitly close where it is not defined. Thus, Eq.~(\ref{dirac}) simply
states that plane waves diffracted by the dipole in the $x$
direction are reflected by the surface with a factor given by the
Fresnel reflection coefficient.

Because $q R(q)$ is an odd function of $q$:
\begin{equation}
E_z(x)=\frac{p_0}{2\pi \epsilon_0}\int_{0}^{+\infty} dq\,q\,R(q)
\sin{qx}. \label{chptot}\end{equation} This integral cannot be
expressed in closed form but can be computed by conventional
numerical techniques. In Fig.\,\ref{fig8} are compared the $z$
components of the electric field computed with the groove
simulation (for $t=w=100$ nm) and with Eq.~(\ref{chptot}). The two
curves, after proper normalization, agree very well. It might
appear surprising that the phase of the dipolar model does not
need to be adjusted compared to the groove calculation, but
Fig.\,\ref{ampphase} shows that the phase of the wave diffracted
by the groove is precisely $\pi$ for this groove geometry. The
overall conclusion is that it is the accumulation of oscillating
charges at the corners of the groove, rather than details of the
groove profile itself, that plays a key role in the global shape
of the diffracted wave a few hundreds of nanometers away from the
groove center.  The field is essentially the field diffracted by a
dipole placed near the surface, its structure is determined by the
fact that the source has a broad-band spatial frequency spectrum,
and that the surface supports a long-lived, guided SPP mode.

An interesting point is the role of constituent propagative and
evanescent modes in the creation of the surface wave. In
particular the composite diffracted evanescent wave (CDEW)
model\,\cite{LT04}, previously invoked to interpret similar
phenomena, considers explicitly only the evanescent part. The two
contributions are difficult to extract from full numerical
simulations but can be easily carried out with the dipole
approach. The scattered field of Eq.~(\ref{chptot}) is separated
in its propagative and evanescent components:
\begin{equation}\label{somme}
E_z(x)=E^{pr}_z(x)+E^{ev}_z(x),
\end{equation}
with:
\begin{eqnarray}
E^{pr}_z(x)&=&\frac{p_0}{2\pi \epsilon_0}\int_0^{k_0} dq\,q\,R(q) \sin{qx}\nonumber\\
E^{ev}_z(x)&=&\frac{p_0}{2\pi \epsilon_0}\int_{k_0}^{+\infty} dq\,q\,R(q) \sin{qx}.\nonumber
\end{eqnarray}
\begin{figure}[htb]
\centering\includegraphics[width=8cm]{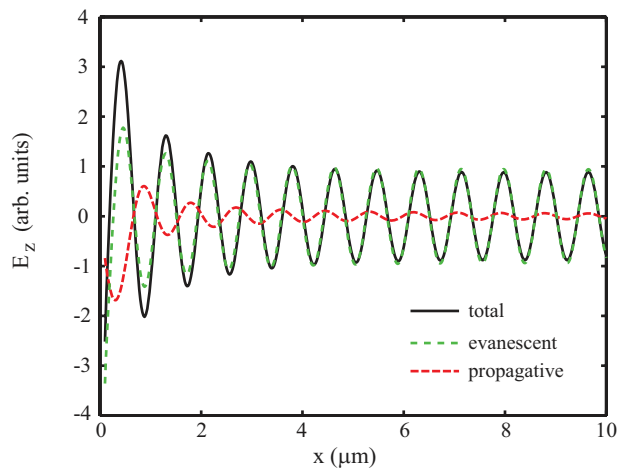}
\caption{Contribution to the field of the radiative and evanescent
components to the total wave calculated from Eq.\,\ref{somme}
(color online).} \label{fig9}\end{figure} Figure \ref{fig9}
compares the real part of these two contributions. The propagative
term represents a substantial fraction, its amplitude being around
50\% of the total wave a few hundreds of nanometers from the
dipole location. However, the amplitude of the propagative
component decreases much faster than the evanescent term; and the
wavelength of the propagative part is clearly longer than the
wavelength of the evanescent component. The reason is that for all
evanescent modes, including the SPP mode, $q>k_0$,
$\lambda <\lambda_0$, and $n_{\mathrm{eff}}>1$. For the
propagative modes $q\leq k_0$, $\lambda\geq \lambda_0$,
$n_{\mathrm{eff}}\leq 1$.

\begin{figure}[htb]
\centering\includegraphics[width=8cm]{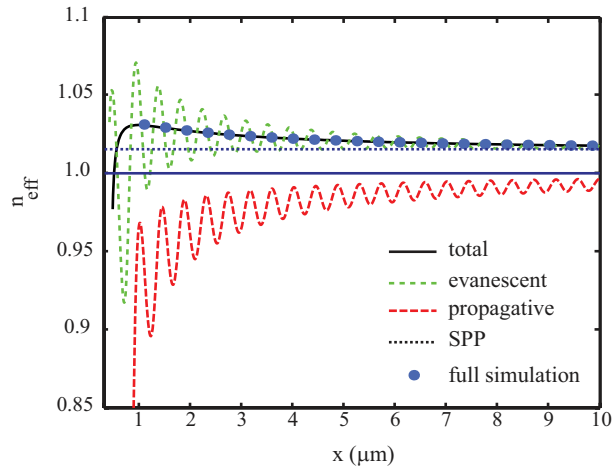}
\caption{Contribution to the effective index of the radiative and
evanescent components to the total wave, and comparison with the
effective index calculated with the groove full simulation (color
online).} \label{fig10}\end{figure}

These trends appear clearly in Fig.\,\ref{fig10}, which represents
the evolution with distance from the dipole of the effective
indices of refraction for the different contributions. The index
of the evanescent part is initially larger than the SPP index and decreases
to this value asymptotically.  The index of the propagative
part, initially less than unity, approaches $n=1$ from below. This is a
consequence of the fact that the propagative wave is dominated by
grazing plane waves, for which the reflection coefficient is
almost equal to 1, implying $q\simeq k_0$ and therefore $n\simeq1$. When
these two contributions are summed, the effective index follows
the curve computed from the groove simulation and is in good
agreement with measurement\,\cite{GAV06a}.

\section{Analytical model}
\begin{figure}
\vspace{0.5cm}
\includegraphics[width=8cm]{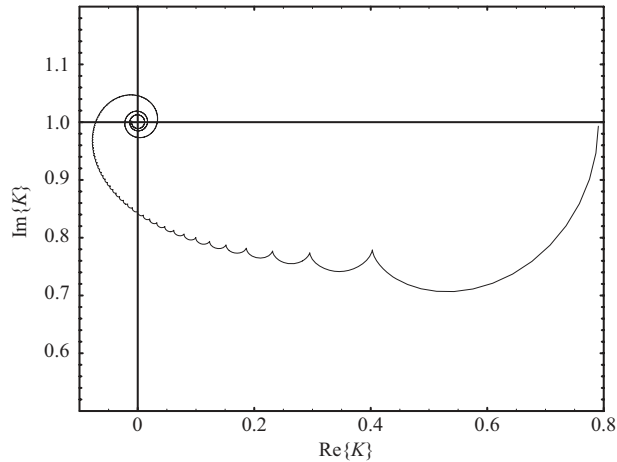}
\caption{\label{fig4} Parametric curve representing the function
$i-\mbox{Ei}[i(1-\alpha)t]/2\pi+\mbox{Ei}[-i(1+\alpha)t]/2\pi$ in
the complex plane. In this example, $\alpha=q_{\mathrm{spp}}/k_0=1.02$. }
\end{figure}

\begin{figure}[ht]
\begin{center}
\includegraphics[width=8cm]{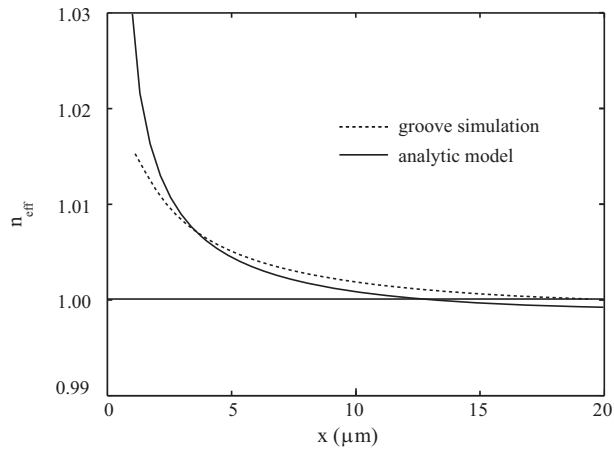}
\caption{\label{figmod}Comparison between the $E_z$, effective
index and the effective index computed from the analytical model.
The groove's dimensions are: $t=100$ nm, $w=100$ nm.}
\end{center}
\end{figure}

The dipole-on-surface model can be further simplified in order to
express the diffracted wave in closed form. In the following, we
present a simple opaque-screen analytic model, similar to that of
Kowarz\,\cite{K95}, but in which the SPP wave is introduced as a
boundary condition on the surface.  In the Kowarz model the
evanescent part of the diffracted wave is computed assuming the
presence of a slit in an infinitely thin opaque screen, and taking
into account only the evanescent components of the field just
above the metal. With these assumptions, the amplitude of the
evanescent wave along the interface is given by:
\begin{eqnarray}
\lefteqn{E(x)=E_0\int_{|q|>|k_0|}^\infty \frac{\sin(q w/2)}{q} e^{i q x} dq \label{eq1}}\\
&&=-\frac{E_0}{\pi}\left[\mbox{Si}\left(k_0(x+\frac{w}{2})\right)-\mbox{Si}\left(
k_0(x-\frac{w}{2})\right)\right],\mbox{ }|x|>w/2,\nonumber
\end{eqnarray}
where Si is the sine integral defined as: $$
\mbox{Si}(\alpha)=\int_0^\alpha \frac{\sin t}{t}\,dt.$$  The
interpretation is straightforward: the slit diffracts the incident
wave into a sum of evanescent waves of spatial frequency $q$ whose
amplitudes are weighted by the Fourier amplitudes of the slit. The
Fourier spectrum of the slit is a cardinal sine (sinc) function.
Moreover, when $x\rightarrow\infty$, the solution of Eq.~\ref{eq1}
is correctly approximated by:
\begin{equation}
E(x)\approx \frac{E_0}{\pi}\frac{d}{x}\cos\left(k_0
x+\frac{\pi}{2}\right).\nonumber
\end{equation}
This model however does not reproduce correctly the result of
Ref.\,\cite{GAV06a}. One of the reasons is that the finite
conductivity of the screen is not included in the Kowarz approach.
If we consider again a TM--polarized wave incident on the groove,
the SPP mode of complex wave vector $q_{\mathrm{spp}}$ is excited
along the interface.  In that case, the corresponding wave vector
is created by the diffraction of the incident wave. The dispersion
relation of the SPP guided mode, $q_{\mathrm{spp}}=k_0
\sqrt{\epsilon/(\epsilon+1)}$, can be retrieved by calculating the
pole of the reflection coefficient of the metal/free-space
interface $R(q)$. At $\lambda_0=852$ nm and value of $\epsilon$
for silver, $q_{\mathrm{spp}}=(1.0154+i\,5.54\,10^{-4})k_0$. For an
evanescent wave whose wave vector is near $q_{\mathrm{spp}}$, the
reflection coefficient can be approximated by
$A/(q-q_{\mathrm{spp}})$, where $A$ is a constant. We can say that
the incident wave impinging on the groove is at first
\emph{diffracted} with amplitude corresponding to the Fourier
spectrum of the groove, and is then \emph{reflected} along the
metallic interface with a coefficient $R(q)$. Hence, an
approximate expression of the evanescent wave propagating along
the interface is obtained by replacing Eq.~(\ref{eq1}) with:
\begin{eqnarray}
 \lefteqn{E(x)=E_0\int_{-\infty}^{-k_0} \frac{\sin(q w/2)}{q} \frac{1}{q+q_{\mathrm{spp}}}e^{i q x} dq }\nonumber\\
 &&+E_0\int_{+k_0}^\infty \frac{\sin(q w/2)}{q} \frac{1}{q-q_{\mathrm{spp}}}e^{i q x} dq. \label{eq2}
\end{eqnarray} Here two poles must be inserted because the SPP wave is excited in both $\pm x$ directions.
This expression can be simplified using the fact that the width of
the groove Fourier spectrum is of the order of $1/w$, whereas the
width of the ``spectral line" of the Fourier SPP mode is of the
order of $\Im(q_{\mathrm{spp}})$, a thousand times narrower than
the Fourier spectrum of the groove. Hence, in $q-$space the groove
structure is essentially a constant over the width of the SPP
response. Changing the width of the groove will only modify the
amplitude of the plane waves of wave vector $q\approx
q_{\mathrm{spp}}$, and thus the amplitude of the diffracted wave.
Hence we have:
\begin{equation}
E(x)\approx E_0\int_{-\infty}^{-k_0} \frac{1}{q+q_{\mathrm{spp}}}e^{i q x} dq+E_0\int_{+k_0}^\infty \frac{1}{q-q_{\mathrm{spp}}}e^{i q x} dq.
\end{equation}
For $x>w/2$, this expression reads:
\begin{equation}
E(x)=2\pi\,E_0 \,e^{i q_{\mathrm{spp}} x}\; K(x),
\end{equation}
with:
\begin{equation}
K(x)=i-\frac{\mbox{Ei}\left(i(k_0-q_{\mathrm{spp}})x\right)}{2\pi}+\frac{\mbox{Ei}\left(i(-k_0-q_{\mathrm{spp}})x\right)}{2\pi}.\nonumber
\end{equation}
The function Ei is called exponential integral and is defined by:
\begin{equation}
\mbox{Ei}(z)=-\int_{-z}^{\infty} \frac{e^{-t}}{t} dt. \nonumber
\end{equation}
It appears that the amplitude of the SPP wave propagating along
$x$, $\exp(i q_{\mathrm{spp}} x)$, is multiplied  by an envelope of complex
value $K(x)$. This function is represented in the complex plane
for typical parameters of SPP wave vector in Fig.~\ref{fig4}(a).
The low values of $x$ correspond to the right of the curve. When
$x$ goes toward infinity, $K(x)$ whirls toward $z_{lim}=i$. The
strong oscillation at the beginning of the curve is due to a
beating between the Ei$(-q_{\mathrm{spp}})$ and the Ei$(+q_{\mathrm{spp}})$ term.
As this function has a varying phase, it will affect the
wavelength of the surface wave. Figure \ref{figmod} compares the
evolution of the surface index for $x$-- and $z$--components using
the index computed from the previous formula.
This analytical model predicts the same trends as the numerical and the dipole
approach: the effective index of the wave generated near the
groove is greater than $n_{\mathrm{spp}}$ but decreases and
converges toward the expected SPP value within a range of about
10~$\mu$m. In fact, the effective index oscillates slightly around
$n_{\mathrm{spp}}$ at larger distances. There is a good
qualitative agreement with the dipole model, because the
evanescent waves play the main role in the creation of the surface
wave. The fact that the reflection coefficient is replaced by a
simple pole in $q_{\mathrm{spp}}$ modifies the time evolution of
the wave amplitude (not shown), but the wavelength evolution is
unchanged. The effective index is overestimated in the first
micrometers near the groove because the radiative part is not taken into account.

\section{Summary}
We have studied in detail the structure of the wave diffracted by
a groove or slit milled in a metallic surface, illuminated by a
monochromatic plane wave. First, the Green's tensor method has been used
 to analyze the amplitude, phase and frequency behavior of the
surface wave in the vicinity of the groove. In this zone, the
surface wave has a transient regime characterized by a rapid
variation of the amplitude within the first 2 to 3 micrometers,
and an increase of the surface wavelength up to the value of the
SPP-wavelength in the first 15 micrometers. The phase and the
amplitude of the scattered wave depends strongly on the groove
geometry, as the incident wave excites an ``organ-pipe" mode
inside the groove.  Best agreement with the experimental results
of Gay et al.~\cite{GAV06a} is obtained assuming a somewhat deeper
groove (120 nm) as in the experiment (100 nm). This value is
within the uncertainty of the actual milled depth using focused
ion beam (FIB) fabrication.   We have also presented a
simplified model in which the surface wave is excited by a
 line dipole parallel to and just above a flat surface without
groove. This approach permits the extraction of an analytical
expression for the $z$ component of the electric field along the
interface. The agreement between this model and the full
simulation is very good, showing that the transient ``near-zone"
regime does not depend on the precise shape of the groove. Indeed
the details of groove depth and width only influences the
amplitude and the phase of the generated wave. The overall form
results from a line dipole source with a broad
$q$ spectrum interacting with a surface that supports a mode.
Moreover, we have studied the influence on the wave
structure of the propagative and the evanescent contributions. The
propagative waves contribute importantly in the first few
micrometers from the source, but their amplitude decay is faster
and their wavelength is longer than the evanescent contribution.
The wavelength of the propagative contribution decreases with the
distance down to the excitation wavelength, whereas the effective
wavelength of the evanescent contribution increases up to the SPP
effective wavelength. Finally, we have studied a simplified model
of the diffraction process, in which the reflection coefficient is
replaced by a pole located at the SPP wave vector in $q$ space.
The scattered field can be then expressed in closed form. This
``minimal model" correctly reproduces the SPP excitation and the
wavelength evolution with distance. Such a semi-analytical model may be useful
for the design and optimization of more elaborate photonic
circuits whose behavior in large part will be controlled by
surface waves.

\section{Acknowledgments}
G.L. and O.J.F.M. acknowledge funding from the IST Network of
Excellence Plasmonanodevices (FP6-2002-IST-1-507879).  J.W.
acknowledges Support from the Minist\`ere d\'el\'egu\'e \`a
l'Enseignement sup\'erieur et \`a la Recherche under the programme
ACI-``Nanosciences-Nanotechnologies," the R\'egion
Midi-Pyr\'en\'ees [SFC/CR 02/22], and FASTNet
[HPRN-CT-2002-00304]\,EU Research Training Network.

\section*{APPENDIX}

The purpose of this appendix is to show location of the dipole,
just above or below the metal/free-space interface is independent
of the results obtained in section \ref{secdip}. We begin with
Eq.~\ref{chptot}:
\begin{equation}
E_z(x)=\frac{p_0}{2\pi k_0^2}\int_{0}^{+\infty} dq\,q\left[R(q)-R(\infty)+R(\infty)\right]\sin{qx}\label{eqap1}
\end{equation}
The integrand diverges when $q\rightarrow +\infty$. However, the
integral is defined for $x>0$. We can write write
Eq.~(\ref{eqap1}) as:
\begin{equation}
E_z(x)=\frac{p_0}{2\pi k_0^2}\left\lbrace\int dq\,q
\tilde{R}(q)\sin{qx}+R(\infty)\int dq\,q \sin{qx},\right\rbrace
\label{eqap2}
\end{equation}
where
\begin{equation}
R(\infty)=\frac{1-\epsilon(\omega)}{1+\epsilon(\omega)}\quad\mbox{and}\quad\tilde{R}\equiv
R(q)-R(\infty)
\end{equation}
so that from Eq.\,\ref{R-de-q}:
\begin{equation}
\tilde{R}(q)=\frac{2\epsilon(\omega)}{1+\epsilon(\omega)}\frac{\sqrt{\epsilon(\omega)k_0^2-q^2}-\sqrt{k_0^2-q^2}}{\sqrt{\epsilon(\omega)k_0^2-q^2}+\epsilon(\omega)\sqrt{k_0^2-q^2}}\nonumber
\end{equation}
The first part of Eq.~\ref{eqap2} converges because:
$$q \tilde{R}(q) \xrightarrow[q\rightarrow +\infty]{ }\frac{\gamma}{q},\quad\mbox{with}\quad\gamma\in \mathbf{C}.$$
The second part is equal to the derivative of a Dirac function. Hence:
\begin{equation}
E_z(x)=\frac{p_0}{2\pi k_0^2} \int dq\,q \tilde{R}(q)\sin{qx}+\alpha \delta'(x)
\end{equation}
The right term of the sum is only a term located at $x=0$. For numerical integration, it is more convenient to use this last expression.

If the dipole is located just under the interface:
\begin{equation}
E_z^-(x)=\frac{p_0}{2\pi k_0^2}\int_{0}^{+\infty} dq\,q T(q) \sin{qx},\nonumber
\end{equation}
with:
\begin{equation}
T(q)=\frac{2\sqrt{\epsilon(\omega)k_0^2-q^2}}{\sqrt{\epsilon(\omega)k_0^2-q^2}+\epsilon(\omega)\sqrt{k_0^2-q^2}}.\nonumber
\end{equation}
Here,
\begin{equation}
T(q)\xrightarrow[q\rightarrow +\infty]{ }T(\infty)=\frac{2}{1+\epsilon(\omega)}.\nonumber
\end{equation}
Hence:
\begin{eqnarray}
E_z^-(x)&=&\frac{p_0}{2\pi k_0^2}\left\lbrace\int_{0}^{+\infty} dq\,q\, \tilde{T}(q)\sin{qx}\right.\nonumber\\
&&\left. \quad\qquad\qquad +T(\infty)\int dq\,q \sin{qx}\right\rbrace\nonumber\\
&=&\frac{p_0}{2\pi k_0^2}\int dq\,q \,\tilde{T}(q)\sin{qx}+\beta\,\delta'(x)
\end{eqnarray}
with
\begin{equation}
\tilde{T}(q)=\tilde{R}(q)\nonumber
\end{equation}
Hence the field diffracted by two dipoles located just above or just under the vaccum/metal interface differs only in $x= 0$.

\end{document}